\documentclass[copyright,creativecommons]{eptcs}
\usepackage{breakurl}
\usepackage{amsmath}
\usepackage{wrapfig}

\def\titlerunning{Ensuring Liveness Properties of Distributed Systems}
\def\authorrunning{R.J. van Glabbeek}

\title{\titlerunning\newline\Large (A Research Agenda)}
\author{Rob van Glabbeek
\institute{NICTA\thanks{NICTA is funded by the Australian Government through the Department of
   Communications and the Australian Research Council through the ICT
   Centre of Excellence Program.}\,    , Sydney, Australia}
\institute{School of Computer Science and Engineering,
University of New South Wales, Sydney, Australia}
\email{rvg@cs.stanford.edu}
}
\providecommand{\publicationstatus}{\Large POSITION PAPER\quad March 3, 2016}
\begin{document}
\maketitle

\begin{abstract}
Often fairness assumptions need to be made in order to establish
liveness properties of distributed systems, but in many situations
these lead to false conclusions.

This document presents a research agenda aiming at laying the
foundations of a theory of concurrency that is equipped to ensure
liveness properties of distributed systems without making fairness
assumptions.  This theory will encompass process algebra, temporal
logic and semantic models, as well as treatments of real-time.  The
agenda also includes developing a methodology that allows successful
application of this theory to the specification, analysis and
verification of realistic distributed systems, including routing
protocols for wireless networks.

Contemporary process algebras and temporal logics fail to make
distinctions between systems of which one has a crucial liveness
property and the other does not, at least when assuming \emph{justness},
a strong progress property, but not assuming fairness.
Setting up an alternative framework involves giving up on identifying
strongly bisimilar systems, inventing new induction principles,
developing new axiomatic bases for process algebras and new congruence
formats for operational semantics, and creating new treatments of time
and probability.

Even simple systems like fair schedulers or mutual exclusion protocols
cannot be accurately specified in standard process algebras (or Petri
nets) in the absence of fairness assumptions.  Hence the work involves
the study of adequate language or model extensions, and their
expressive power.
\end{abstract}

\section{State-of-the-art and objectives}

\subsection*{Specification, analysis and verification of distributed systems}

At an increasing rate, humanity is creating distributed systems
through hardware and software---systems consisting of multiple
components that interact with each other through message passing or
other synchronisation mechanisms.  Examples are distributed databases,
communication networks, operating systems, industrial control systems,
etc. Many of these systems are hard to understand, yet vitally
important. Therefore, significant effort needs to be made to
ensure their correct working.

Formal methods are an indispensable tool towards that end.
They consist of specification formalisms to unambiguously capture the
intended requirements and behaviour of a system under consideration, tools and
analysis methods to study and reason about vital properties of the
system, and mathematically rigorous methods to verify that (a) a
system specification ensures the required properties, and (b) an
implementation meets the specification.

The standard alternative to formal specification formalisms are descriptions
in English, or other natural languages, that try to specify the
requirements and intended workings of a system. History has shown,
almost without exception, that such descriptions are riddled with
ambiguities, contradictions and under-specification. 
Formalisation of such a description---regardless in which
formalism---is the key to elimination of these holes.

A formal specification of a distributed system typically comes in (at
least) two parts.

One part formulates the \emph{requirements}
imposed on the system as a list of properties the system should
have. Amongst the formalisms to specify such requirements are temporal
logics like Linear-time Temporal Logic (LTL) \cite{Pn77} or Computation Tree Logic
(CTL) \cite{EC82}.  Amongst others, they can specify \emph{safety properties},
saying that something bad will never happen, and \emph{liveness properties},
saying that something good will happen eventually \cite{Lam77}.

The other part is a formal description of how the system ought to
work on an \emph{operational} (= step by step) basis, but abstracting from
implementation details. For distributed systems such accounts
typically consist of descriptions of each of the parallel components,
as well as of the communication interfaces that specify how different
components interact with each other. Languages for giving such formal
descriptions are \emph{system description languages}.  When a
system description language features constants to specify elementary
system activities, and operators (like parallel or sequential
composition) to create more complex systems out of simpler ones, it is
sometimes called a \emph{process algebra}.
Alternatively, operational system descriptions can be rendered in a
model of concurrency, such as Petri nets or labelled transition systems.
Such models are also used to describe the meaning of system
description languages.

Once such a two-tiered formalisation of a system has been provided,
there are two obvious tasks to ensure the correct working of
implementations: (a) guaranteeing that the operational system description
meets the requirements imposed on the system, and (b) ensuring that an
implementation satisfies the specification. The latter task additionally
requires a definition of what it means for an implementation to
satisfy a specification, and this definition should ensure that
any relevant correctness properties that are shown to hold for the
specification also hold for the implementation.

A third type of task is the study of other properties of the
implementation, not implied by the specification.  Examples are
measuring its execution times, when these are not part of the
specification, or its success rate, for operations for which success
cannot be guaranteed and only a best effort is made. Potentially,
these tasks call for applications of probability theory.

Traditional approaches to ensure the correct working of distributed
systems are simulation and test-bed experiments. While these are
important and valid methods for system evaluation, in particular for
quantitative performance evaluation, they have limitations in regards
to the evaluation of basic correctness properties. Experimental
evaluation is resource-intensive and time-consuming, and, even after a
very long time of evaluation, only a finite set of operational
scenarios can be considered---no general guarantee can be given about
correct system behaviour for a wide range of unpredictable deployment
scenarios.
I believe that formal methods help in this regard; they complement
simulation and test-bed experiments as methods for system evaluation
and verification, and provide stronger and more general assurances
about system properties and behaviour.

\subsection*{Achievements of process algebra and related formalisms}

\emph{Process algebra} is a family of approaches to the specification,
analysis and verification of distributed systems. Its tools encompass
algebraic languages for the specification of processes (mentioned
above), algebraic laws to reason about processes,
and induction principles to derive behaviours of infinite systems from
those of their finite approximations.

Many industrial size distributed systems have been successfully
specified, analysed and verified in frameworks based on process algebra.
Examples can be found though the following links.
Major toolsets primarily based on process algebra include
\href{http://www.cs.ox.ac.uk/projects/fdr/}{FDR} \cite{GRABR14},
\href{http://cadp.inria.fr/}{CADP} \cite{GLMS11},
\href{http://www.win.tue.nl/mcrl2/}{mCRL2} \cite{GM14} and
the \href{http://www.it.uu.se/research/group/mobility/applied/psiworkbench}{Psi-Calculi Workbench} \cite{BGRV15,BJPV11}.
Most of these toolsets also use model checking or other mathematical
techniques that explore the state spaces of distributed systems.
Similar toolsets primarily based on the latter techniques include
\href{http://spinroot.com/spin/}{SPIN} \cite{SPIN},
\href{http://www.uppaal.org/}{UPPAAL} \cite{BDL04},
\href{http://www.prismmodelchecker.org/}{PRISM} \cite{KNP10} and
\href{http://research.microsoft.com/en-us/um/people/lamport/tla/toolbox.html}{TLA} \cite{La02}.

\subsection*{Verification of routing protocols for wireless networks}

In \cite{FGHMPT12a,TR13,BrGH16} my colleagues and I have developed a process algebra that is tailored to
wireless networks---it features novel treatments of local broadcast,
conditional unicast, and data handling---and used it to study AODV
\cite{rfc3561}---a popular routing protocol for wireless networks,
standardised by the IETF\@. Established methods turned out to be
sufficient for proving safety properties of this protocol, in
particular \emph{loop freedom} \cite{GHPT16}. However, standard techniques turned
out to be fundamentally inadequate to establish \emph{route discovery}
or \emph{packet delivery} \cite{TR13}, two fundamental liveness properties of
routing protocols. Part of my research agenda is to lay
the theoretical foundations to accurately formulate and prove such
properties, and use them to continue my work on network protocols.
I will use this case study to illustrate a few points made below.

\subsection*{Liveness, fairness assumptions, and their dangers}

One of the crucial tasks in the analysis of distributed systems is the
verification of liveness properties, saying that something good will
happen eventually. A typical example is the verification of a communication
protocol---such as the \emph{alternating bit protocol} \cite{Lyn68,BSW69}---that
ensures that a stream of messages is relayed correctly, without loss or
reordering, from a sender to a receiver, while using an unreliable
communication channel.  The protocol works by means of acknowledgements,
and resending of messages for which no acknowledgement is received.

Naturally, no protocol is able to ensure such a thing, unless we
assume that attempts to transmit a message over the unreliable channel
will not fail in perpetuity. Such an assumption, essentially saying
that if one keeps trying something forever it will eventually succeed, is
often called a \emph{fairness} assumption.

Making a fairness assumption is indispensable when verifying the
alternating bit protocol. If one refuses to make such an assumption,
no such protocol can be found correct, and one misses a change to
check the protocol logic against possible flaws that have nothing to
do with a perpetual failure of the communication channel.

For this reason, fairness assumptions are made in many process-algebraic
verification methods, and are deeply ingrained in their methodology \cite{BBK87a}.
This applies, amongst others, to the default incarnations of the
process algebras CCS \cite{Mi90}, ACP \cite{BW90,Fok00,BBR10}, the
$\pi$-calculus \cite{Mi99,SW01} and mCRL2 \cite{GM14}.

Using a fairness assumption, however, needs to be done with care.
Making a fairness assumption can lead to patently false results.
This applies for instance to the communication protocol
in cases where one of the possible behaviours of the unreliable
channel is to perpetually loose all messages.
A \emph{global} fairness assumption, as commonly used in
process algebra, leads to false conclusions on packet delivery for routing protocols.

In defence of making a fairness assumption it is sometimes argued that
whenever at some point the probability of success is 0, the success
possibility should not be part of our model of reality. When
infinitely many choices remain that allow success with a fixed positive
probability, with probability 1 success will be achieved eventually.
This argument rests on assumptions on relative probabilities of
certain choices, but is applied to models that abstract from those probabilities.
My counter-argument is that (1) when abstracting from probabilities
it is quite well possible that a success probability is always
positive, yet quickly diminishing, so that the cumulative success
probability is less than 1, and (2) that in many applications we do
not know whether certain behaviours have a chance of occurring or
not, but they are included in the model nevertheless.

\subsection*{Process algebras without fairness assumptions, and their limitations}

In certain process algebras, such as CSP \cite{Ho85,Ros97}, it is not
common to make any fairness assumptions. There also are variants of
CCS, ACP, etc.\ that do not make them \cite{Wa90,GLT09b}.

Here I will argue that virtually all these approaches \emph{cannot}
make sufficiently strong progress assumptions to establish meaningful
liveness properties in realistic applications.
Namely, I will show two programs, $P$ and $Q$, that are equated in
virtually all process-algebraic approaches to date. In technical
terms, they are \emph{strongly bisimilar} \cite{Mi90,vG00}. Yet, there
is a crucial liveness property that holds for $P$ but not for $Q$, when not
assuming fairness. So the process algebra must either claim that both
programs have the liveness property, which in case of $Q$ could be an
unwarranted conclusion, possibly leading to the design of systems with
dangerous or catastrophic behaviour, or it falls short in asserting the
liveness property of $P$.
\\[1ex]
\mbox{}\hfill
$x:= 1\quad \| \quad\textbf{repeat}~~ y:=y+1 ~~\textbf{forever}$
\hfill($P$)
\\[1ex]
Program $P$ is the parallel composition of two non-interacting processes,
one of which sets the variable $x$ to $1$, and the other repeatedly
increments a variable $y$. I assume that both variables $x$ and
$y$ are initialised to $0$.
\begin{wrapfigure}[5]{r}{5cm}
\vspace{-1.5ex}
\begin{tabbing}\qquad\=\textbf{repeat}\\
\> \quad \= \textbf{case} \\
\>\> \quad \= \textbf{if} ~\texttt{True}~ \textbf{then}~ y:=y+1 ~\textbf{fi}\=\\
\>\>\> \textbf{if} ~$x=0$~ \textbf{then}~ x:=1 ~\textbf{fi}\\
\>\> \textbf{end} \\
\>\textbf{forever}\>\>\>(Q)
\end{tabbing}
\end{wrapfigure}
In program $Q$ the \emph{case}-statement is interpreted such that if the
conditions of multiple cases hold, a non-deterministic choice is made
which one to execute.
The programs $P$ and $Q$ are strongly bisimilar; both can be
represented by means of the following labelled transition system:
\vspace{-5pt}

\expandafter\ifx\csname graph\endcsname\relax
   \csname newbox\expandafter\endcsname\csname graph\endcsname
\fi
\ifx\graphtemp\undefined
  \csname newdimen\endcsname\graphtemp
\fi
\expandafter\setbox\csname graph\endcsname
 =\vtop{\vskip 0pt\hbox{%
    \special{pn 8}%
    \special{ar 2000 450 50 50 0 6.28319}%
    \special{sh 1.000}%
    \special{pn 1}%
    \special{pa 2118 353}%
    \special{pa 2035 415}%
    \special{pa 2079 321}%
    \special{pa 2118 353}%
    \special{fp}%
    \special{pn 8}%
    \special{pa 2035 485}%
    \special{pa 2250 750}%
    \special{pa 2550 750}%
    \special{pa 2550 150}%
    \special{pa 2250 150}%
    \special{pa 2042 407}%
    \special{sp}%
    \graphtemp=.5ex
    \advance\graphtemp by 0.450in
    \rlap{\kern 2.930in\lower\graphtemp\hbox to 0pt{\hss $y:=y+1$\hss}}%
    \special{ar 1000 450 50 50 0 6.28319}%
    \special{sh 1.000}%
    \special{pn 1}%
    \special{pa 1025 300}%
    \special{pa 1000 400}%
    \special{pa 975 300}%
    \special{pa 1025 300}%
    \special{fp}%
    \special{pn 8}%
    \special{pa 1000 0}%
    \special{pa 1000 300}%
    \special{fp}%
    \special{sh 1.000}%
    \special{pn 1}%
    \special{pa 1850 425}%
    \special{pa 1950 450}%
    \special{pa 1850 475}%
    \special{pa 1850 425}%
    \special{fp}%
    \special{pn 8}%
    \special{pa 1050 450}%
    \special{pa 1850 450}%
    \special{fp}%
    \graphtemp=\baselineskip
    \multiply\graphtemp by -1
    \divide\graphtemp by 2
    \advance\graphtemp by .5ex
    \advance\graphtemp by 0.450in
    \rlap{\kern 1.500in\lower\graphtemp\hbox to 0pt{\hss $x:=1$\hss}}%
    \special{sh 1.000}%
    \special{pn 1}%
    \special{pa 904 331}%
    \special{pa 965 415}%
    \special{pa 872 369}%
    \special{pa 904 331}%
    \special{fp}%
    \special{pn 8}%
    \special{pa 965 485}%
    \special{pa 650 750}%
    \special{pa 350 750}%
    \special{pa 350 150}%
    \special{pa 650 150}%
    \special{pa 957 408}%
    \special{sp}%
    \graphtemp=.5ex
    \advance\graphtemp by 0.450in
    \rlap{\kern 0.000in\lower\graphtemp\hbox to 0pt{\hss $y:=y+1$\hss}}%
    \hbox{\vrule depth0.717in width0pt height 0pt}%
    \kern 2.930in
  }%
}%

\mbox{}\qquad\;\;{\box\graph}
\vspace{2ex}

As a warm-up exercise, one may ask whether the variable $y$ in $P$ or
$Q$ will ever
reach the value $7$---a liveness property. A priori, I cannot give a
positive answer, for one can imagine that after incrementing $y$ three
times, the program for no apparent reason stops making
progress and does not get around to any further activity. In most
applications, however, it is safe to assume that this scenario will
not occur. To accurately describe the intended behaviour of $P$ or $Q$,
or any other program, one makes a \emph{progress assumption},
saying that if a program is in a state where further activity is
possible (and this activity is not contingent on input from the
environment that might fail to occur) some activity will in fact
happen. This assumption is sufficient to ensure that in $P$ or $Q$ the
variable $y$ will at some point reach the value $7$.

Progress assumptions are commonplace in process algebra and many other
formalisms. They are explicitly or implicitly made in CCS, ACP, the
$\pi$-calculus, CSP, etc., whenever such formalisms are employed to
establish liveness properties. Temporal logics, such as LTL
\cite{Pn77} and CTL \cite{EC82}, have progress assumptions built in,
and can formalise the statement that $y$ will\ in fact reach the value $7$.

A more interesting question is whether $x$ will ever reach the value $1$.
This liveness property is \emph{not} guaranteed by progress
assumptions as made in any of the standard process algebras or temporal logics.
The problem is that all these formalisms rest on a model of
concurrency where parallel composition is modelled as arbitrary interleaving.
The programs $P$ and $Q$ have computations like
$$\begin{array}{lllllll}
x:=1;& y:=y+1;& y:=y+1;& y:=y+1;& y:=y+1;& y:=y+1;& \dots \\
y:=y+1;& x:=1;& y:=y+1;& y:=y+1;& y:=y+1;& y:=y+1;& \dots \\
y:=y+1;& y:=y+1;& y:=y+1;& y:=y+1;& x:=1;& y:=y+1;& \dots \\
\end{array}$$
where the action $x:=1$ can be scheduled arbitrary far in the sequence
of $y$-incrementations, but also a computation
\begin{equation}\tag{$C^\infty$}
\begin{array}{lllllll}
y:=y+1;& y:=y+1;& y:=y+1;& y:=y+1;& y:=y+1;& y:=y+1;& \dots \\
\end{array}
\end{equation}
in which $x:=1$ never happens, because $y:=y+1$ is always scheduled instead.
For this reason, temporal logic as well as process algebra---when not
making fairness assumptions---say that $x$ is not guaranteed to reach
the value $1$, regardless whether talking about $P$ or $Q$.

When assuming that parallel composition is implemented by
means of a scheduler that arbitrarily interleaves actions from both
processes, this conclusion for $P$ appears plausible. However, when $\|$
denotes a true parallel composition, where the program $P$ consists
of two completely independent processes, it appears more reasonable
to use a stronger progress assumption that does guarantee that $x$
will reach the value $1$. Such a strong progress assumption is
formalised in the context of process algebra in my paper
\cite{GH15a}, written jointly with Peter H\"ofner, and called \emph{justness}.
We also applied it in the study of routing protocols for wireless networks \cite{TR13}.

However, whereas my justness assumption disqualifies the computation
$C^\infty$ for $P$, it is entirely reasonable to allow it for $Q$.
After all, the \textbf{case}-statement may be implemented in such a
way as to always pick the first case that applies.
A fairness assumption would (unjustly) eliminate this computation even
for $Q$; here I argue for a theory of concurrency in which such a fairness
assumption is not made.

Hence, virtually all existing process-algebraic approaches equate
two programs of which one has the liveness property that eventually
$x$ will reach the value $1$, and the other does not, at least not
without assuming fairness. So those approaches that do not assume
fairness lack the power to establish this property for $P$.

\subsection*{The same limitations of temporal logics}

At first sight, justness is exactly the same as what is called
\emph{justice} by Lehmann, Pnueli \& Stavi \cite{LPS81}. In fact, their
motivating example for the proposal of justice is similar to
program $P$. They called a computation \emph{just}
``if it is finite or if every transition which is continuously enabled
beyond a certain point is taken infinitely many times.''
This proposal disqualifies the computation without the assignment
$x:=1$ for the program $P$.

To determine whether it also disqualifies this computation for the
program $Q$ it matters how one formalises the notion of a transition
being ``continuously enabled''. 
The literature following \cite{LPS81} has systematically interpreted this
as meaning ``in every state'' (``beyond a certain point'').
The computation $C^\infty$ of $Q$ that only has transitions $y:=y+1$ has the
property that in each of its states (that is between two such transitions)
the transition $x:=1$ is enabled. For this reason, that computation
would be disqualified by the notion of justice attributed to \cite{LPS81}.
This formalisation of the concept of justice from \cite{LPS81} is
commonly known as \emph{weak fairness}.

An alternative formalisation of a transition being ``continuously enabled'',
based on the principle of ``noninstantaneous readiness'' of
\cite{AFK88}, postulates that the enabledness of the transition $x:=1$
is interrupted when execution the conflicting transition $y:=y+1$,
even though it is again enabled in the state following this transition.
In \cite{GH15a} we have formalised this interpretation of
enabledness of transitions, and shown that the resulting notion of
justice coincides with our own notion of justness.

Yet, since most work on temporal logic identifies the notion of
justice from \cite{LPS81} with a concept of fairness that makes no
difference between the programs $P$ and $Q$, it appears prudent to use
a subtly different name for the concept of justness.  The
conclusion is that in standard approaches to temporal logic, just like
in process algebra, either a concept of fairness is used that makes
dangerous predictions---i.e.\ that in $Q$ the variable $x$ will
eventually reach the value $1$---or no such concept of fairness is
used, resulting in the inability to derive important liveness
properties---such as that in $P$ the variable $x$ will eventually
reach the value $1$.
The notions of route discovery and packet delivery for routing
protocols are among those liveness properties.

\subsection*{Goal}

This brings me to my research agenda in this matter: the development
of a theory of concurrency that is equipped to ensure liveness
properties of distributed systems, incorporating justness assumptions
as explained above, but without making fairness assumptions.  This
theory will encompass process algebra, temporal logic, Petri nets and
other semantic models, as well as treatments of real-time, and of the
interaction of probabilistic and nondeterministic choice.

Since this involves distinguishing programs that are strongly
bisimilar, it requires a complete overhaul of all the basic
machinery that has been built in the last few decennia.
It requires new equivalence relations between processes, new
axiomatisations, new induction principles to reason about infinite
processes, new congruence formats for operational semantics ensuring
compositionality of operators, and new extensions with time and
probabilities.

As in the absence of fairness assumptions some crucial systems like
fair schedulers or mutual exclusion protocols cannot be accurately
specified in Petri nets or standard process algebras \cite{GH15b}, it
also involves the study of adequate model or language extensions, and
their expressive power.

My agenda furthermore aims at developing a methodology that allows successful
application of the envisioned theory of concurrency to the specification, analysis and
verification of realistic distributed systems, focusing on cases where
the new balance in establishing liveness properties bears fruit.
In particular, I advocate the application of this work to the analysis of routing protocols in wireless
networks; using the new theory of concurrency I would like to extend our
prior work \cite{TR13,FGHMPT12b,HGTPMF12,GHTP13,GHPT16,BrGH16}, amongst others by formally
proving route discovery and package delivery properties of suitable protocols.
\vspace{5pt}

\noindent
This research agenda involves the following tasks:
\vspace{-1pt}\leftmargini 20pt
\begin{enumerate}\parskip 0pt\itemsep 0pt
\item To formally define what it means for a process algebra to
  encompass justness but not fairness.
\item To investigate and classify semantic equivalences (necessarily
  incomparable with strong bisimilarity) that respect liveness
  when assuming justness.
\item To study liveness and justness properties in non-interleaved
  semantic models like Petri nets, event structures and higher
  dimensional automata.
\item To find complete axiomatisations and adequate induction
  principles for process algebras with justness.
\item To find syntactic requirements that guarantee that the relevant
  justness-preserving equivalences are congruences for operators
  specified conform those requirements.
\item To study the necessary extensions to process algebras or Petri nets
  to model simple processes like fair schedulers, and investigate the
  relative expressiveness of process algebras with and without them.
\item To re-evaluate the possibility and impossibility results for
  encoding synchrony in asynchrony when insisting that justness
  properties are preserved.
\item To extend relevant justness preserving formalisms with treatments
  of real-time.
\item To adapt the existing testing theory for nondeterministic
  probabilistic processes to a setting where justness is preserved.
\item To apply the obtained formalisms to the further study of routing
  protocols for wireless networks, paying special attention to liveness
  properties like package delivery.
\end{enumerate}

\section{Detailed research agenda}
\newcounter{wp}
\newcommand{\workp}[1]{\subsection*{Task \thewp\refstepcounter{wp}: #1}}

Although stable foundations of process algebra and related
verification methods have been laid in the last 30 years, surprisingly
little of it can be reused as foundation for a theory, as needed here,
that encompasses justness but not fairness. Much of the existing work
is predicated on the belief that at least strongly bisimilar processes
need not be distinguished, and as it is necessary to give this up,
a totally new way of thinking on the relevant foundations is in order.

Below I describe the main tasks that I see as part of this process.

\refstepcounter{wp}
\workp{Process algebra and temporal logic}

The first task is to formulate a precise definition of what it means
for a process-algebraic specification and verification formalism to
encompass justness but not fairness. Moreover, although I want to
avoid a \emph{global} fairness assumption, applying to all choices of
represented systems, there needs to be a way to specify \emph{local}
fairness assumptions, ones that can be justified for particular
scheduling problems in the systems under consideration, and for those only.

My proposals in this matter have already been formulated in
\cite{GH15a}---combining process algebra and temporal logic to tackle
local fairness properties---but that paper is specific to the process algebra CCS
and its extension with broadcast communication. In \cite{TR13} we have
indicated how the same concepts apply to the process algebra AWN,
tailored to wireless networks.

What is lacking so far is a general theory saying how to do this for arbitrary
process algebras, possibly involving a CSP-style communication
mechanism \cite{Ho85,Ros97}, priorities \cite{CLN01}, or name-binding \cite{Mi99,SW01}.

\workp{A classification of semantic equivalences and preorders}

A crucial prerequisite for verifying that an implementation meets a
specification is a definition of what this means. Such a definition
can be given in the form of an equivalence relation or preorder on
a space of models that can be used to describe both specifications and
implementations. For sequential systems, an overview of suitable
preorders and equivalence relations defined on labelled transition
systems is given in \cite{vG01,vG93}. Preorders and equivalences
specifically tailored to preserve safety and liveness properties are
explored in \cite{vG10}. Equivalences for non-sequential systems are
discussed, e.g., in \cite{GG01}. They include \emph{interleaving equivalences},
in which parallelism is equated with arbitrary interleaving, as well
as equivalence notions that take, to some degree, concurrency
explicitly into account. In \cite{vG15} I show that none of these
equivalences respect \emph{inevitability} \cite{MOP89}: when assuming
justness but not fairness, they all equate systems of which only one
has the property that all its runs reach a specific success state.
Hence, none of these equivalences are suitable for a process-algebraic
framework destined to establish liveness properties under the justness
assumption advocated above.

As shown in \cite{vG10}, safety and liveness properties are intimately
linked with the notions of may- and must-testing of De Nicola \&
Hennessy \cite{DH84}. However, \cite{vG10} also treats
\emph{conditional liveness} properties that surpass the power of
must-testing. In \cite{vG09} I proposed a notion of \emph{reward testing},
and claimed that it exactly matches with conditional liveness properties.

My first goal in this task is to substantiate the above claim.
Once that is done, the same testing framework can be applied to derive
preorders for concurrent processes that respects (conditional) liveness
properties in the presence of the justness assumption.
This may yield a result similar to the fair failure preorder of \cite{Vo02}.
Additionally, variants of most existing preorders and equivalences may
be found that take respect liveness under justness assumptions.

Possibly, just forcing an existing preorder to respect liveness by
adding appropriate clauses to its definition gives a result that is
less suitable for verification tasks. It will for instance most likely fail to
satisfy the \emph{recursive specification principle} of \cite{BK86,BBK87a},
saying that guarded recursive specifications have unique solutions.
That problem might be addressed by formulating more discriminating
preorders and equivalences that do not feature explicit conditions on
infinite runs, yet respect liveness properties.
A first proposal for such an equivalence is the \emph{structure
  preserving bisimilarity} of \cite{vG15}. That equivalence is
probably too discriminating for many verification tasks, so more
research is called for.

\workp{Petri nets and other semantic models}

The standard semantics of process algebras is in terms of labelled
transition systems. However, for accurately capturing causalities
between event occurrences, models like Petri nets, event structures or
higher dimensional automata are sometimes preferable. As shown above,
unaugmented labelled transition systems are not sufficient to capture
liveness properties when assuming justness but not fairness.  On the
other hand, Petri nets offer a structural characterisation of
justness: if a transition is enabled, and none of the tokens enabling
it are ever consumed by a competing transition, then it will
eventually fire. For this reason, interpretations of process algebras
in terms of these models \cite{GM84,Wi84,GV87,Wi87a,Old91,GV03,vG06}
need to be studied in relation to formulations of the relevant
semantic equivalences in terms of these models.

\workp{Complete axiomatisations and induction principles}

Many process-algebraic verifications \cite{Ba90} employ principles
like the above-mentioned \emph{recursive specification principle}
and the \emph{approximation induction principle} \cite{BK86,BBK87a},
allowing to derive properties of infinite systems through analysis of
their finite approximations. It is likely that these principles do not
hold in straightforward variants of existing semantic equivalences
that respect liveness when assuming justness. The two ways to cope
with that are (1) searching for finer equivalences that do not have
this shortcoming, or (2) searching for alternative induction principles that
hold and are useful in verification.
As this time I cannot say which of these directions is the most
promising; more research is in order here.

Algebraic laws have also shown their use in verification, and the
isolation of a complete collection of such laws is often the starting
point of both a good verification toolset and a better understanding
of the semantic concepts involved. For these reasons, finding complete
axiomatisations of suitable equivalences to deal with justness and
liveness is an important task.

\workp{Congruence formats for structural operational semantics}

In process-algebraic verification it is essential that composition
operators on processes, such as the parallel composition, are
\emph{compositional} w.r.t.\ the semantic equivalence employed.
This means that the composition of two processes, each given as an
equivalence class of, say, labelled transition systems, is independent
on the choice of representatives within these equivalence classes.
Compositionality of an operator w.r.t.\ an equivalence is the same as
the equivalence being a congruence w.r.t.\ the operator.

Starting with \cite{dS85,BIM95,GrV92}, the most elegant and efficient
way to establish compositionality results in process algebra is by
means of \emph{congruence formats}, sets of syntactic restrictions
on the operational specification of the behaviour of composition
operators that ensure compositionality. This line of research is
continued in \cite{Gr93,BolG96,Ul92,Ver95,vG04,FG96,Bl95,Ul00,UP02,UY00,Fok00b,BFG04,FGW06,vG11,GMR06,FGW11}.

One of the tasks on my agenda is the search for congruence formats
tailored to the equivalences produced by Task 2.

\workp{Expressiveness}

In the absence of fairness assumptions even simple systems like fair
schedulers or mutual exclusion protocols cannot be accurately
specified in standard process algebras or Petri nets.
This is shown in \cite{Vo02,KW97,GH15b}.
For this reason, my research agenda involves the study of adequate
extensions to standard process algebras, as well as to models of
concurrency like Petri nets, event structures and higher dimensional automata.
A plausible extension to process algebras to fill this gap are signals
\cite{Be88b}; alternatives are broadcast communication \cite{GH15a} or
priorities \cite{GH15b}. A plausible extension to Petri are read arcs \cite{Vo02}.
The relative expressiveness of these models and languages needs to be
studied, along with other arguments for one model or language over another.

The resulting study on the relative expressiveness of process
algebras ought to be placed within the formal frameworks developed for
comparisons of expressive power provided in \cite{Bo85,Gorla10a,vG12}.

In \cite{dS85,vG94a} results are obtained saying that all languages
with a structural operational semantics of a certain form can be
expressed in versions of the process algebras {\sc Meije} \cite{AB84}
and ACP \cite{BK86a}, respectively. Ideally, such results are to be
obtained also for the language extensions contemplated above.
This may involve a standardisable mechanism akin to structural
operational semantics for specifying justness aspects of process
combinators, in addition to their step-by-step behaviour.

\workp{Asynchronous interaction in distributed systems}

During the last few years I have been involved in joint research with
Ursula Goltz (TU Braunschweig) and Uwe Nestmann (TU Berlin) aiming to
determine to what extent synchronous communication can be simulated by
asynchronous communication. This research was done in models like
Petri nets as well as in terms of process algebras with mobility
primitives, like the $\pi$-calculus. Building on earlier work
\cite{Nes00,NP00,Pal03,CCP07}, we established separation
results, saying that in some conditions synchronous communication
cannot be mimicked by asynchronous communication, and encodability results,
showing how in other situations it could be done
\cite{GGS08b,GGS08d,SPG11,PSN11,PN12,GGS13,PNG13,MSG14}.
All those results are dependent on the choice of a semantic
equivalence, to judge whether a proposed encoding of synchronous into
asynchronous communication has been successful.

Hence these results need to be revisited when using semantic
equivalences with the appropriate respect for justness.
It would be interesting to see if this changes the relations found so
far, and if so, to what extent.

\workp{Real-time}

Extensions of process algebras and semantic models with treatments of
time are vital for many applications. Although many timed process
algebras already exists \cite{RR88,HR95,Wang90,MT90,MT91,BB91,NS94,BM02,OS06},
it is unclear whether any of them could form a natural extension of
an untimed process algebra with justness as envisioned in this project.

As an example of a process that is hard to encode in some of the
proposals above, consider the parallel composition of two components
that each await a communication from the environment and then
synchronise with each other.  The problematic part is that although,
at the synchronisation point, either component can wait arbitrary long
on the other, the synchronisation between the two will
happen as soon as both components are ready for it, and does not admit
further lingering after that point.

The goal of this task is to find a suitable process algebra that
provides a timed extension for a suitable untimed proposal with justness,
and to do a formal comparison of the expressiveness of the existing
process algebras for timed systems. The latter can be done by mapping
them all into a formalism that includes the expressive power of each
of the process algebras to be compared.

In the setting of wireless network protocols, a step towards the first
goal is \cite{BrGH16}.

\workp{Extensions with probabilistic choice}

In collaboration with Yuxin Deng, Matthew Hennessy, Carroll Morgan and
Chenyi Zhang \cite{DGHMZ07,DGMZ07,DGHM08,DGHM09,DGHM13}, building on
earlier work by Wang Yi, Kim Larsen, Bengt Jonsson, Chris Ho-Stuart
and Roberto Segala \cite{WL92,JHW94,Seg96}, I have developed a theory
of testing for processes with nondeterminism and probabilistic choice.
This theory determines the coarsest semantic equivalences for
probabilistic processes that respect safety and liveness properties.
A natural task is to extend this work to a parallel processes with
justness, thereby extending the non-probabilistic work in Task 2.

\workp{Network protocols}

The final task on my agenda is the application of the obtained framework
to the specification, analysis and verification of routing protocols. This involves:
\begin{itemize}
\item enriching our process algebra of wireless networks (AWN \cite{FGHMPT12a}) with
  time, and use it to specify a timed version of the AODV routing
  protocol (based on the existing untimed formalisation of AODV \cite{TR13,GHPT16});
  this is work to appear as \cite{BrGH16};
\item formally specifying a number of similar protocols that are used in industry
  and standardisation bodies, including AODVv2 \cite{AODVv2}, HWMP
  \cite{IEEE80211s}, OLSR \cite{rfc3626}, OLSRv2 \cite{rfc7181} and BATMAN
  \cite{batman}---the latter three heavily depend on time and need a
  timed version of AWN;
\item establishing safety properties like loop freedom for some of these
  protocols, or show why such properties do not hold;
\item establishing liveness properties like route discovery or packet
  delivery for some of these protocols, or show why such properties do
  not hold;
\item proposing modifications of these protocols that have the required
  safety and liveness properties, or design a new protocol based on
  similar ideas if the latter is hopeless;
\item extending the above work to the link layer of the protocol
  stack, by formally specifying the CMSA protocol \cite{ieee802.11} and studying its
  relevant properties.
\end{itemize}
Besides this, other applications of the obtained framework to the
specification, analysis and verification of realistic distributed
systems are worthy of study as well.

\section{Conclusion}

I have presented a research agenda aiming at laying the foundations of
a theory of concurrency that is equipped to ensure liveness properties
of distributed systems without making fairness assumptions.  The
agenda also includes the application of this theory to the specification,
analysis and verification of realistic distributed systems.
I have divided the detailed agenda into 10 tasks, each of which in
some sense involves solving one or more open problems.

It is not my intension to address all tasks listed here myself---there
are simply too many tasks for that to work out. I hope that this
document stimulates others to do some work in this area.

\bibliographystyle{eptcsalphaini}
\bibliography{agenda}
\end{document}